\documentclass{sig-alternate-2013}

\newfont{\mycrnotice}{ptmr8t at 7pt}
\newfont{\myconfname}{ptmri8t at 7pt}
\let\crnotice\mycrnotice%
\let\confname\myconfname%

\usepackage{balance}  
\usepackage{graphics} 
\usepackage{times}    
\usepackage{url}      

\makeatletter
\def\url@leostyle{%
  \@ifundefined{selectfont}{\def\UrlFont{\sf}}{\def\UrlFont{\small\bf\ttfamily}}}
\makeatother
\def\pprw{8.5in}
\def\pprh{11in}

\usepackage[pdftex]{hyperref}
\hypersetup{
pdftitle={SIGCHI Conference Proceedings Format},
pdfauthor={LaTeX},
pdfkeywords={SIGCHI, proceedings, archival format},
bookmarksnumbered,
pdfstartview={FitH},
colorlinks,
citecolor=black,
filecolor=black,
linkcolor=black,
urlcolor=black,
breaklinks=true,
}

\begin{document}

\title{Tag Me Maybe:\\Perceptions of Public Targeted Sharing on Facebook}

\numberofauthors{1}
\author{
  \alignauthor Saiph Savage ~ Andr\'{e}s Monroy-Hern\'{a}ndez ~ Kasturi Bhattacharjee ~ Tobias H{\"o}llerer\\
\affaddr{University of California Santa Barbara} \\
\affaddr{Universidad Nacional Aut\'{o}noma de M\'{e}xico} \\
\affaddr{Microsoft Research}
\email{\{saiph, kbhattacharjee, holl \}@cs.ucsb.edu saiph@unam.mx amh@microsoft.com}
}

\maketitle
\begin{abstract}
Social network sites allow users to publicly tag people in their posts. These tagged posts allow users to share to both the general public and a targeted audience, dynamically assembled via notifications that alert the people mentioned. We investigate people's perceptions of this mixed sharing mode through a qualitative study with $120$ participants. We found that individuals like this sharing modality as they believe it strengthens their relationships. Individuals also report using tags to have more control of Facebook's ranking algorithm, and to expose one another to novel information and people. This work helps us understand people's complex relationships with the algorithms that mediate their interactions with each another. We conclude by discussing the design implications of these findings.
\end{abstract}

{\bf Author Keywords} Social media; access controls; social networks; narrowcast; broadcasting; algorithmic filtering\\\\
{\bf ACM Classification Keywords} {H.5.3 Group and Organization Interfaces}

\section{Introduction}
On social networking sites (SNS), people have contacts from different facets of their lives, e.g., college or work. This can lead to unintentionally sharing sensitive  content with subsets of friends~\cite{hogan2010presentation}. People engage in a spectrum of sharing modes to overcome this problem, from targeted sharing, where messages are shared with specific individuals, e.g., in an email~\cite{bersteinFeedme}, to public broadcasts where people share messages that are appropriate for all~\cite{hogan2010presentation}. 

There has been growing interest in understanding people's perceptions of these sharing modalities. Bernstein {et al.}~\cite{bersteinFeedme} studied targeted sharing in private messages and found that people saw this modality as a way to share personally relevant content. 
\begin{figure}[t]
\centering
\includegraphics[width =0.30\textwidth]{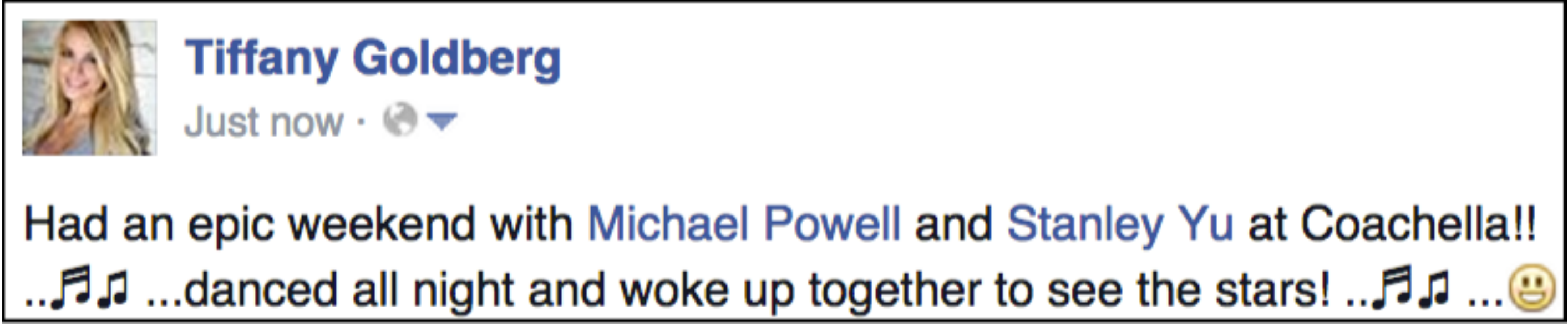}
\caption{\small Example of public targeted sharing on Facebook. 
}
\label{fig:samplePost}
\end{figure}
Kairam {et al}.~\cite{talkingCircles} used Google+ to study selective sharing, i.e., the sharing of content with specific predefined groups such as  ``family'' or ``work.'' Their results indicate that people use selective sharing to evangelize, ask questions, or start conversations.  Hogan~\cite{hogan2010presentation}, in studies of public broadcasting in social media, found that in this sharing mode, people often speak to the ``lowest common denominator'' to allow everyone to understand the message. 

Few studies have fully addressed the public targeted sharing modality of SNS. Figure~\ref{fig:samplePost} presents an example of a Facebook post with this sharing mode. Public targeted sharing combines elements of targeted sharing and broadcasting. When a person tags others, the tagged individuals  receive a direct notice of the post. However, the post's public nature also reaches a wider audience similar to a broadcast (e.g., the entire social graph of the content producer). This hybrid sharing mode also represents a public display of social connections, as the public posts carry the names of certain friends and links to their personal Facebook profiles~\cite{Donath:2004:PDC:1031314.1031348}. Previous research has studied particular aspects of this interaction, such as the identity concerns it triggers~\cite{Lampinen:2011:TIM:1978942.1979420} or the types of conversations that such posts elicit~\cite{huang2010conversational,10.1109/HICSS.2009.602}. Little is known,  however, regarding people's different views and uses of this sharing modality.

The aim of this investigation is to organize and understand the variety of  perceptions of this hybrid sharing mode on Facebook. Although Facebook has a particular design for public targeted sharing, the concepts involved are not unique. Sites such as Twitter or Soundcloud also provide hyperlinks to the profiles of the people tagged in content. Google+ links tags to people's profiles and allows people to  broadcast to the social circles of those tagged. Gold users of reddit receive notifications when others tag them.

Our interview-based study of 120 individuals discovers that individuals feel public targeted sharing strengthens friendships.  They also believe it can be used to overcome the algorithmic filtering powers in  play, bringing each other greater exposure to more surprising content and people. 

\section{Methods}
To examine the perceptions of public targeted sharing, we employ a methodology similar to that used in a study about people's perceptions of Facebook disclosures~\cite{Lampinen:2011:TIM:1978942.1979420}. Similar to Lampinen et al.~\cite{Lampinen:2011:TIM:1978942.1979420}, instead of pursuing a longitudinal study, our goal is to shed light on the variety of interpretations that individuals have regarding this sharing modality. We conducted interviews to understand the perceptions. 
\newpage
\subsection{Data Collection} 
We recruited participants both offline and online to avoid having only highly active Facebook users. To recruit online, we posted invitations to our study in different Facebook groups and to pages randomly selected from public listings. 
To recruit offline, we approached people in public spaces such as bus stops and parks and invited them to take part in our study. We labeled participants recruited directly from Facebook as  F\# and those recruited from public spaces as P\#.  Our recruitment material avoided the term ``tagging'' to not exclude people who rarely tag.  

Of the 120 individuals participated in our study, 32 were recruited from Facebook, and 88 from public spaces. Participants used Facebook to varying degrees. The individuals recruited from Facebook were 18-52 years old;  44\% were female, 56\% were male. The participants from public spaces were 18-68 years old; 52\% were female, and 48\% were male. The interviews had a median duration of 11 minutes, with the longest lasting 40 minutes, and the shortest 8 minutes. The interviews covered participants' interpretations of this sharing modality from the perspectives of the different social roles involved, such as content producers (individuals who tag friends in content they share on Facebook), taggees (individuals tagged in  the content), and viewers (individuals who view the tagged content).  We counterbalanced the order in which we asked participants about each role. Participants voiced which roles they had taken and only responded for those that they had personally assumed.
All participants had been viewers at least once, 87\% had been taggees at least once, and 83\% had been content producers at least once. 

Our study began by asking each participant to write down examples of public targeted sharing. On average, each participant provided two examples (for instance, two Facebook posts where friends were tagged). For each role, we used the examples to start a discussion. For instance, we asked: ``What are your thoughts when someone tags you in this type of post?'' We encouraged the participants to lead the discussion. Our goal was mainly to get them to elaborate and exemplify.  Participants mostly reported on their perceptions of public targeted sharing in posts (Facebook status updates or wall posts). Some also referred to tagging in comments, photos, or videos. However, participants did not highlight any difference among content types. The interviews also covered participants' background in SNS use (sites and devices used, type of content they or their friends share, number of Facebook  friends, etc.) We did not witness any difference in interview responses across age groups or SNS usage. 

\subsection{Categorization Perspectives} 

We used qualitative coding to analyze interview responses. This allowed us to  establish a categorization of perceptions of public targeted sharing. Our overarching goal was to identify patterns in these perceptions. We transcribed all audio recordings and separated each interview into sections based on the social role that the interviewee was reporting on. Two researchers then jointly read each interview transcript for each social role to identify key concepts and ideas. These initial concepts were then discussed with a third researcher.  We decided that a category would cover a general type of perception and that, when possible, it would include perspectives specific to each social role. Concepts were then aggregated and similar ones renamed. 

We also quantified the number of participants who viewed the tagging of Facebook users in posts as a mechanism that involved two mixed audiences. We considered this would help to further contextualize our results. Two researchers, unfamiliar with the work, read each interview and classified whether the person saw tagging as involving two audiences. The researchers found that all interviewees referred to  tagging as a mixed sharing mode. Additionally, 71\% of the interviewees specifically used the term ``share'' to describe tagging on Facebook. 
\section{Results}
The following five main categories cover participants' perceptions of public targeted sharing.  One person can be associated with more than one category, as a person can have more than one perception associated with the sharing modality. 
We did not identify differences between participants' Facebook usage and their perceptions on sharing.

{\bf Stronger Relationships.} A majority of our interviewees (79\%) perceived public targeted sharing as a way to build stronger relationships.  From our interviews, it appears that tagging builds stronger friendships because the targeted audience feels that someone is making time for them:

\begin{quotation}
~\emph{``I think the best thing about being tagged is that there is somebody who is considering and thinking about you. I always comment on the things people tag me because they took the time to think about me.''~28, male F2.}
\end{quotation}

This result matches theory on disclosure, which states that the more an interaction makes people feel singled out, the more value they attribute to it~\cite{Donath:2004:PDC:1031314.1031348}. Being publicly tagged has become a social signal that helps  individuals feel unique and noteworthy. Content producers appear to be aware of this dynamic. Interviewees stated that, via public targeted sharing, they secured higher-quality responses from their network: 
\begin{quotation}
~\emph{``[Tagging] ensures that you are working with only loyal people that you know will help and promote your cause, not sabotage it.''~30, male, F1.}
\end{quotation}
Individuals appear to use targeted sharing to jump start discussions with sympathetic crowds. Interviewees also believed that, via tagging, Facebook's ranking algorithm was more likely to share their content with others. Therefore some people use public targeted sharing to revive relationships:
\begin{quotation}
\emph{``...I basically hope that friends I haven't seen in a while will see this [a tagged post], and they will comment, and we can reconnect [...] Basically I have had other people in mind when I have tagged -- sort of thinking about the fact that they are also watching.''~21, male, F55.}
\end{quotation}

{\bf Surprising Content.}  Interviewees  (61\%) perceived  this sharing mode as a way to expose audiences to information outside of what they normally viewed on Facebook. Interviewees believe that friends have similar tastes. Facebook's recommendation algorithm thus tends to show more of one type of content to the members of a particular social circle. Tagging broke this by enabling content producers to reach multiple social graphs  and share surprising information, outside the social graph's norms: 
\begin{quotation}
\emph{``I like to tag people that I know are interested in something and whose audience will also care. But my interest is in creating a crossover. So I involve audiences that follow people in dance, but  I share with them something totally different, such as poetry...''~25, female, F10.} 
\end{quotation}
This is similar to the finding of Huang {et al.} \cite{huang2010conversational} of how people tagged on Twitter to direct content to certain social graphs. However, perhaps due to Facebook's recommendation algorithms, we observed here an emphasis on using tagging to diversify a social graph's information. 

Participants also had folk theories about how tags influenced Facebook's ranking algorithm:
\begin{quotation}
~\emph{``If I tag someone and then he comments, the post becomes ``active'' and it'll appear at the top of the News Feed. So I'll sometimes tag someone directly in the post, or in the comments. In either case, if it makes him comment, the post will be pushed to the top.''~25, female, F10.} 
\end{quotation}
Participants also believed that this sharing mode facilitates the introduction of audiences to new and surprising people:
\begin{quotation}
~\emph{ ``Being a group of friends, I want them to meet other people who they might find interesting. The whole idea of social media is that we can share. So they might see a face, but they don't know who they are; with the tag, they can find more information about them''~26, male, P3.} 
\end{quotation}

{\bf Public Image.} Interviewees (60\% ) believed that public targeted sharing can help refine one's public image:
\begin{quotation}
~\emph{``I tag friends to represent something I want to project. So let's say that I want to project that I am a super hipster, well then I will tag friends in support of a cause, I will casually mention that my friends and I  are out writing in our Moleskine journals...''~29, female F5.}
\end{quotation}

Interviewees also reported that they tagged to link their content to the reputation or image of their friends and increase the value of the content.  Participants cherry-picked who they tagged. 
They analyzed how influential or relevant the image of a person whom they planned on tagging was in reference to the content they wanted to share. 
Similarly, participants also believed that {Facebook's ranking algorithm} would give more visibility to posts linked to high-profile users. Individuals thus tagged users with influential public images in an attempt to garner more visibility for the posts they wanted to broadcast:
\begin{quotation}
~\emph{``Facebook will always rank higher content tied to verified, well known accounts. You're losing a lot of visibility if you don't link your content. You have to select well who you tag.''~41, male, P83. }
\end{quotation}
In contrast to selective sharing~\cite{talkingCircles}, public targeted sharing is also a public display of social connections~\cite{Donath:2004:PDC:1031314.1031348}. This might cause individuals to relate it to public image construction, as everyone (people and algorithms included) can view how one interacts with others.

Participants saw public targeted sharing as an opportunity to tailor not only their own public image but also that of their friends. They tagged as a way to collectively help friends find content that favored them. Interviewees also expressed that, in general, they felt honored to be tagged and be a part of their friends' posts.  However, some also voiced identity concerns because others selected the content linked to them and the content might not {match} what they wanted to {portray}:
\begin{quotation}
\emph{``[Tagging] is a very public thing. It allows anyone to broadcast something about you, something that might mislead the idea of who you are. Through tags your identity can be created for you. Your identity becomes what other people broadcast about you, their idea of you.''~26, male, O3.}
\end{quotation}
Content producers were aware that others are self-conscious about the type of content that is posted about them. We found that most empathized with their friends and usually alerted them offline that they were going to tag them:
\begin{quotation}
~\emph{``A lot of people don't want to be tagged at a party. People are self-conscious of what they want or what they don't want about them online. [...] I always ask people if they want to be tagged. I am in the car and I say:
Hey guys I am making a Facebook status and I am going to tag all you guys!''~19, male O7}.
\end{quotation}

This echoes the findings of Lampinen {et al.}~\cite{Lampinen:2011:TIM:1978942.1979420} regarding people negotiating 
self-disclosures before conflicts emerge. However, we found that, despite the negotiations taking place, individuals occasionally had to untag themselves. 
This occurred when individuals were excessively tagged in their friends' content. Individuals wanted to show their involvement in their friends' lives (as it helped strengthen friendships). Yet, showing too much of that involvement, e.g.,  being tagged in too many posts, obscured who they were:
\begin{quotation}
\emph{ ``In every wedding photo they posted, they included my name regardless of whether I was in the photo or not [...] Because I don't have any other posts on my profile, that one wedding event began dominating my page and it began characterizing me because it was the only thing that people saw on my profile. When people went into my page, they thought I was really into weddings [...] My identity was made up by someone's wedding. I felt that was too emphasized. I thought about untagging myself from the photos but decided to leave them. I felt it was sad to disassociate myself from my friend's wedding.''~29, male, O8.} 
\end{quotation}
The nature of this mixed sharing modality generates struggles, as individuals want to show support for friends, while also portraying their desired public image. 

{\bf Unwanted Content.} Interviewees (38\%) perceived this sharing modality as annoying, because it brought them undesired data. Individuals thought that the content should have been shared privately with the targeted audience. Some taggees also found this sharing modality annoying because they received excessive Facebook notifications. Some untagged  themselves as a result.  Taggees and viewers alike felt annoyed with tagged posts generated by companies, especially if they were advertisements. Individuals appeared tolerant of targeted ads created by friends. They considered that at least their friends thought about them to tag them and that  the posts were a chance to be updated about their friends' lives: 
\begin{quotation}
\emph{ ``It's usually interesting to see what my friends are selling. One girl was tagging so we could see her ceramics creations [...] it's really interesting to see  what they're up to...''~25, female, P52.} 
\end{quotation}
Content producers are aware that this sharing modality can be annoying. They are thus careful about how often they publicly target others and in what content they target others. Interestingly, some see this annoyance as a chance to play with friends:
\begin{quotation}
~\emph{``One friend of mine was an Obama supporter [...] I was tagging them in some anti-Obama stuff to mess with them [...] I succeeded in being annoying...hehe...''~24, male, F15.}
\end{quotation}
Despite this, taggees and viewers felt that unwanted content could also bring them serendipitous discoveries, as it could expose them to interesting strangers:
\begin{quotation}
~\emph{``...I have personally met several people through this ``spam'', like people with whom I've had an interest to collaborate, and people with whom I've shared interests on this and other topics. It isn't common that you get to meet people in spam. They are rare exceptions. Marvelous exceptions that bring you marvelous opportunities.''~20, male, F3.} 
\end{quotation}
{\bf Recollection.} Interviewees (33\%) perceived public targeted sharing as a type of virtual diary that helped to document who played a part in their lives. Participants reported that they enjoyed returning later to these posts to reminisce. Interviewees thought it helped them to reminisce not only individually but also collectively. Individuals surfaced posts from the past to  encourage their audiences to relive memories:
\begin{quotation}
\emph{ ``...For me, they [tagged posts] are more like a path for sharing memories. They are really about making your friends relive these experiences again...''~28, female, P60.}
\end{quotation}
Content producers also have the perception that by including a person's name in a post, they can implicitly remind the taggees of events that they will conduct together. In this case, content producers admitted to mentioning certain persons so that their posts would  be favored by Facebook's ranking algorithm and people would be reminded of their event:
\begin{quotation}
\emph{ ``...tagging helps to build the fan base of your event, and people will be constantly reminded of the event.  Their friends are also more likely to see it [i.e., the content] if you tag. That's one of the things which  Facebook loves and favors: tagging!''~21, male, F202.} 
\end{quotation}
\section{Discussion}
We used Facebook as a medium to investigate perceptions on an increasingly popular sharing modality: public targeted sharing. In general, individuals feel that this modality helps to reinforce relationships, as content producers publicly show that they are taking the time to consider a particular targeted audience. Tagged posts generated automatically by companies are likely to not be perceived as positively, because individuals do not see any real person making time for them. This is similar to what was observed in the Scratch online community, where people valued  credit granted automatically less than credit given by humans~\cite{Monroy-Hernandez:2011:CCG:1978942.1979452}. SNS and UI designers could contemplate how to help companies better engage with their online clients, perhaps by encouraging more humanized sharing. This result is important especially given the recent lawsuit over people's name appearing in tagged advertising. 

Individuals appeared to use public targeted sharing to build collaborative spaces, such as a space to reminisce collectively about the past, or a type of backstage to collectively help friends craft a desired public image. Individuals seem to have adopted public targeted sharing as a building block to create the flexible online social spaces they desire. We believe that it is important to design digital structures that allow people to collectively experiment with building on the technology. It is not about just user-testing all elements of an interface but, rather, testing whether people can make use of those elements to construct jointly the dynamic spaces they want.

Our study revealed that individuals view public targeted sharing as a way to  expose each other to surprising content or to other people beyond those recommended by Facebook's algorithms. Recommendation algorithms in general have sought to filter out opinions, people, and items that are different from us, potentially limiting the diffusion of information. Individuals appear to use this sharing modality as a means to free audiences from these ``algorithmic biases'' and distribute information that they consider fresh and interesting.

Individuals also perceive public targeted sharing as a way to reach audiences outside 
their immediate social circles. Facebook does not officially 
present tagging as a way to reach foreign social graphs but rather as a way simply to let people know when they are involved in the posts that are shared. 
Our findings thus raise the question of whether social media would 
benefit from more official digital structures tailored for targeting and assessing  
novel audiences~\cite{savagevisualizing2, forbes2012visualizing,forbes2010behaviorism, DBLP:conf/cscw/SingerFCTSS13}. Audiences could be bombarded with more unwanted content in this setting, however, it might also enable more serendipitous discoveries. Future work could further explore this trade-off. 

Across categories, we uncovered how individuals feel they can use this sharing modality to try to manipulate or ``game'' Facebook's ranking algorithm and 
obtain the viewership they want. This result is consistent with the 
recent work of Gillespie~\cite{gillespie2013relevance}, which points out, ``Teens have been known to tag their status updates with unrelated brand names, in the hopes that Facebook will privilege those updates in their friends' feeds.'' Editors were traditionally the gatekeepers of information. Today, it is filtering algorithms that determine what is relevant~\cite{Tufekci:2015:AOM:2675133.2697079} and even how we interact with content, products and prices \cite{hannak2014measuring}. Our results raise the question of what the role of human-centered SNS should be regarding the algorithmic powers in play. Again, there is a trade-off: allowing manipulations could expose audiences to more unwanted content, yet it might also help disseminate more diverse, unpredictable, noteworthy content.

Finally, similar to \cite{Lampinen:2011:TIM:1978942.1979420}, we call for caution in generalizing our results. 
Although we took care to recruit a diverse set of participants, the pool of interviewees recruited outdoors belonged to particular cultural settings.  We tried to counter this issue by triangulating their responses with those participants recruited online. However, future work could focus on a quantitative analysis with a more varied population.

\balance
{\bf Acknowledgements} Special thanks to all our participants, as well as Leif Singer, Angus Forbes, Airi Lampinen for the immense feedback and iterations on this work. This work was partially supported by the U.S. Army Research Laboratory under Cooperative
Agreement No. W911NF-09-2-0053 and by NSF grant IIS-0747520.

\bibliographystyle{acm-sigchi}
\bibliography{sample}

\begin{thebibliography}{10}

\bibitem{bersteinFeedme}
Bernstein, M.~S., Marcus, A., Karger, D.~R., and Miller, R.~C.
\newblock Enhancing directed content sharing on the web.
\newblock In {\em Proc. of the SIGCHI Conf. on Human Factors in Computing
  Systems}, CHI '10, ACM (2010), 971--980.

\bibitem{Donath:2004:PDC:1031314.1031348}
Donath, J., and Boyd, D.
\newblock Public displays of connection.
\newblock {\em BT Technology Journal 22}, 4 (Oct. 2004), 71--82.

\bibitem{forbes2010behaviorism}
Forbes, A.~G., Hollerer, T., and Legrady, G.
\newblock behaviorism: A framework for dynamic data visualization.
\newblock {\em Visualization and Computer Graphics, IEEE Transactions on 16}, 6
  (2010), 1164--1171.

\bibitem{forbes2012visualizing}
Forbes, A.~G., Savage, S., and H{\"o}llerer, T.
\newblock Visualizing and verifying directed social queries.
\newblock In {\em IEEE Workshop on Interactive Visual Text Analytics. Seattle,
  WA}, Citeseer (2012).

\bibitem{gillespie2013relevance}
Gillespie, T.
\newblock The relevance of algorithms.
\newblock {\em Media Technologies\/} (2013).

\bibitem{hannak2014measuring}
Hannak, A., Soeller, G., Lazer, D., Mislove, A., and Wilson, C.
\newblock Measuring price discrimination and steering on e-commerce web sites.
\newblock In {\em Proceedings of the 2014 Conference on Internet Measurement
  Conference}, ACM (2014), 305--318.

\bibitem{hogan2010presentation}
Hogan, B.
\newblock The presentation of self in the age of social media.
\newblock {\em Bulletin of Science, Technology \& Society\/} (2010).

\bibitem{10.1109/HICSS.2009.602}
Honeycutt, C., and Herring, S.~C.
\newblock Beyond microblogging: Conversation and collaboration via twitter.
\newblock {\em 2013 46th Hawaii International Conference on System Sciences
  0\/} (2009), 1--10.

\bibitem{huang2010conversational}
Huang, J., Thornton, K.~M., and Efthimiadis, E.~N.
\newblock Conversational tagging in twitter.
\newblock In {\em Proceedings of the 21st ACM conference on Hypertext and
  hypermedia}, ACM (2010), 173--178.

\bibitem{talkingCircles}
Kairam, S., Brzozowski, M., Huffaker, D., and Chi, E.
\newblock Talking in circles: selective sharing in google+.
\newblock In {\em Proc. of the SIGCHI Conf. on Human Factors in Computing
  Systems}, CHI '12, ACM (2012), 1065--1074.

\bibitem{Lampinen:2011:TIM:1978942.1979420}
Lampinen, A., Lehtinen, V., Lehmuskallio, A., and Tamminen, S.
\newblock We're in it together: interpersonal management of disclosure in
  social network services.
\newblock In {\em Proceedings of the SIGCHI Conference on Human Factors in
  Computing Systems}, CHI '11 (2011).

\bibitem{Monroy-Hernandez:2011:CCG:1978942.1979452}
Monroy-Hern\'{a}ndez, A., Hill, B.~M., Gonzalez-Rivero, J., and boyd, d.
\newblock Computers can't give credit: How automatic attribution falls short in
  an online remixing community.
\newblock In {\em Proc. of the SIGCHI Conf. on Human Factors in Computing
  SystemsI} (2011).

\bibitem{savagevisualizing2}
Savage, S., Forbes, A., Toxtli, C., McKenzie, G., Desai, S., and Hollerer, T.
\newblock Visualizing targeted audiences.
\newblock In {\em Proceedings of the International Conference on the Design of
  Cooperative Systems}, COOP'15 (2015).

\bibitem{DBLP:conf/cscw/SingerFCTSS13}
Singer, L., Filho, F. M.~F., Cleary, B., Treude, C., Storey, M.-A.~D., and
  Schneider, K.
\newblock Mutual assessment in the social programmer ecosystem: an empirical
  investigation of developer profile aggregators.
\newblock In {\em Proc. of the 2013 Conf. on Computer-Supported Cooperative
  Work} (2013), 103--116.

\bibitem{Tufekci:2015:AOM:2675133.2697079}
Tufekci, Z.
\newblock Algorithms in our midst: Information, power and choice when software
  is everywhere.
\newblock In {\em Proceedings Conference on Computer Supported Cooperative Work
  \& Social Computing}, CSCW '15 (2015).

\end{thebibliography}
\end{document}